# Giant Non-resonant Infrared Second Order Nonlinearity in γ-NaAsSe$_2$

*Jingyang He,[†] Abishek K. Iyer,[†] Michael J. Waters, Sumanta Sarkar, James M. Rondinelli, Mercouri G. Kanatzidis,\* Venkatraman Gopalan\**

[†] Equal contributions

J. He, Prof. V. Gopalan
Department of Materials Science and Engineering, Pennsylvania State University, University Park, Pennsylvania, 16802, USA
Email: vxg8@psu.edu

Dr. A. Iyer, Dr. S. Sarkar, Prof. M. Kanatzidis
Department of Chemistry, Northwestern University, Evanston, Illinois 60208, USA
Email: m-kanatzidis@northwestern.edu

Dr. M. Waters, Prof. J. Rondinelli
Department of Materials Science and Engineering, Northwestern University, Evanston, Illinois, 60208, USA

Keywords: second harmonic generation, nonlinear optics, infrared nonlinear optical crystals

Infrared laser systems are vital for applications in spectroscopy, communications, and biomedical devices, where infrared nonlinear optical (NLO) crystals are required for broadband frequency down-conversion. Such crystals need to have high non-resonant NLO coefficients, a large bandgap, low absorption coefficient, phase-matchability among other competing demands, e.g., a larger bandgap leads to smaller NLO coefficients. Here, we report the successful growth of single crystals of γ-NaAsSe$_2$ that exhibit a giant second harmonic generation (SHG) susceptibility of $d_{11}$=590 pm V$^{-1}$ at 2μm wavelength; this is ~ eighteen times larger than that of commercial AgGaSe$_2$ while retaining a similar bandgap of ~1.87eV, making it an outstanding candidate for quasi-phase-matched devices utilizing $d_{11}$. In addition, γ-NaAsSe$_2$ is both Type I and Type II phase-matchable, and has a transparency range up to 16μm wavelength. Thus γ-NaAsSe$_2$ is a promising bulk NLO crystal for infrared laser applications.

The past decade has seen considerable interest in new nonlinear optical (NLO) crystals for infrared laser applications.[1–7] NLO crystals can combine or split photons to generate new colors starting from a given laser line.[8,9] They are thus used to produce coherent laser radiation over a broad spectral range from the ultraviolet to 15-20 μm and beyond, which is of great importance in many technologies, such as in medical surgery,[10] environmental monitoring,[11] and imaging devices.[12] Though many NLO materials such as $KTiOPO_4$ (KTP),[13] $\beta$-$BaB_2O_4$,[14] and $LiNbO_3$[15] have been employed for generating light in the visible regime, they are not suitable for the infrared region because of their lower conversion efficiencies and infrared absorption past 4.5-5μm wavelength. Although there are several new highly promising materials emerging from various research laboratories,[16–18] currently only a few infrared NLO materials are commercially available such as $AgGaS_2$,[19] $AgGaSe_2$ [20–22] and $ZnGeP_2$.[23] A central goal of the laser materials community is to develop new NLO crystals to complement and to improve upon the current commercial crystals. This is by no means an easy task since there are many competing demands on NLO crystals: High nonlinear coefficients, large transparency range and hence a large bandgap (which unfortunately scales inversely with nonlinear coefficients), low optical absorption coefficient, phase matchability of the NLO process for high efficiency conversion, the ability to grow large, high quality single crystals, and high laser damage threshold among others.[6,17,18,24]

Metal chalcogenides have attracted attention owing to their excellent optical properties arising from their more polarizable electron cloud and weaker interatomic bonds compared to oxides.[6,7] Among the promising infrared NLO materials, γ-$NaAsSe_2$ (**Figure** 1a) exhibits several attractive properties. Its bandgap value of ~1.87eV is comparable with that of $AgGaSe_2$ (1.80eV),[25] but the averaged SHG response is much greater than that of $AgGaSe_2$, characterized using the Kurtz-Perry Powder technique.[26,27] Although this technique can quickly evaluate a potentially promising material, for practical applications, it is necessary to determine the complete anisotropic linear and nonlinear optical property tensors, which are accessible only in high-quality single crystals. The complete linear and nonlinear optical susceptibility tensors of γ-NaAsSe$_2$ have remained unknown since it was first grown in 2009, and only the bandgap and powder SHG were characterized.[27] Theoretical calculations on γ-NaAsSe$_2$ suggested a $\chi^{(2)}$ = 324.6 pm/V suggesting that this compound has the highest SHG response for materials with bandgap greater than 1.5 eV.[28] One of the challenges in the synthesis of the γ-NaAsSe$_2$ is that it undergoes a phase transition at 450 °C to the centrosymmetric δ-NaAsSe$_2$ upon reheating (Figure S1). In this study, we have successfully overcome this problem to grow large enough γ-NaAsSe$_2$ single crystals for the first time in order to fully characterize their anisotropic linear and NLO properties, something that was not possible earlier with crystals that were too tiny.[27] Single crystals of γ-NaAsSe$_2$ exhibit a giant second harmonic generation (SHG) susceptibility of $d_{11}$=590 pm V$^{-1}$ at 2μm, compared to 33 pm V$^{-1}$ for AgGaSe$_2$ while retaining a similar bandgap of ~1.87eV. This makes it an outstanding candidate for exploring superior quasi-phase-matched devices. Using linear and nonlinear optical characterization combined with density functional theory, we determine that γ-NaAsSe$_2$ is both Type I and Type II phase-matchable. These outstanding properties suggest that γ-NaAsSe$_2$ could be a promising bulk NLO crystal for next-generation infrared laser applications.

γ-NaAsSe$_2$ crystallizes in the non-centrosymmetric monoclinic space group *Pc* and exhibits one dimensional infinite [AsSe$_2$]$^-$ chains connected via AsSe$_3$ units along the *a*-axis. For the first time, single crystals of γ-NaAsSe$_2$ were successfully grown using the melt growth with sizes larger than 1mm$^2$ in dimensions, details of which can be found in the Methods. Single crystal X-ray diffraction (XRD) was performed for structural determination (Figure 1b). The detailed crystallographic data and the atomic coordinates are shown in Table S1 and S2 in the Supporting Information. Successful synthesis of the γ-NaAsSe$_2$ phase was also possible by the Bridgeman method at very slow translation speeds (0.5 mm h$^{-1}$), see the Supporting Information. Since the undesirable δ-NaAsSe$_2$ phase is only obtained upon remelting the γ-NaAsSe$_2$, single crystals of the γ-NaAsSe$_2$ phase were obtained by melt growth (very slow cooling rate of 1.25 °C/h, see Methods) and Bridgeman method (very slow translation speeds 0.5 mm h$^{-1}$, see the Supporting Information).

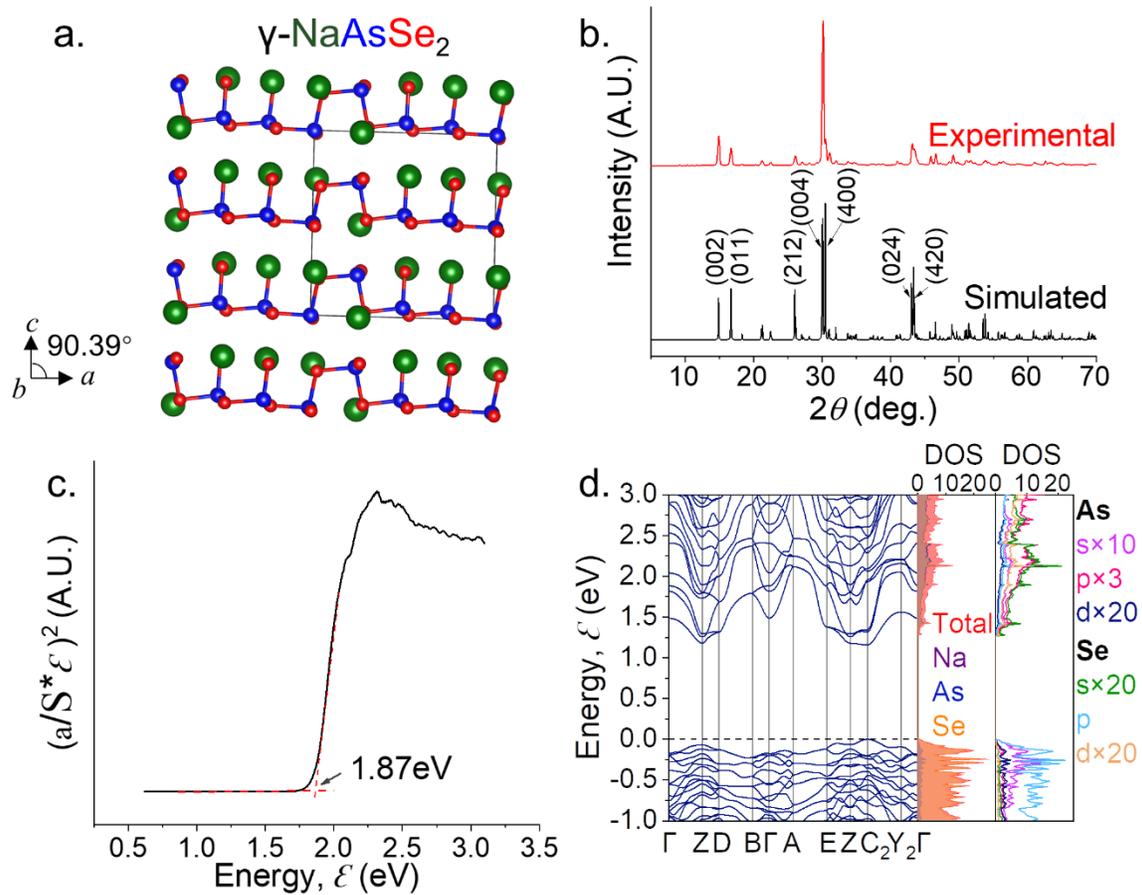

**Figure 1**. (a) Crystal structure of γ-NaAsSe$_2$ viewed along the *b*-axis. (b) Single crystal XRD pattern of γ-NaAsSe$_2$ confirming phase purity. (c) Tauc plot of γ-NaAsSe$_2$ showing the direct bandgap of 1.87eV. (d) Band structure of γ-NaAsSe$_2$ with total DOS and PDOS of As and Se.

Since the bandgap strongly limits both the laser damage threshold (LDT) and the SHG response of a material,[24,29] we characterized the optical transitions of γ-NaAsSe$_2$ using optical diffuse reflectance measurements converted to absorption data using the Kubelka-Munk equation.[30] The electronic band structure calculations on γ-NaAsSe$_2$ indicated a direct bandgap which is consistent with the electronic absorption spectra which shows a value of 1.87 eV, as seen in Figure 1c. The spectra were derived by using the Kubelka-Munk function (α/S) to (α/S)$^2$ for direct bandgap and (α/S)$^{1/2}$ for indirect bandgap (Figure S2, Supporting Information). [30–32] The upper limit of the transparency range of γ-NaAsSe$_2$ was determined by Attenuated Total Reflection Fourier-Transform Infrared Spectroscopy (ATR-FTIR) in the spectral region of 2.5 to 16μm (4000 to 600 cm$^{-1}$), see the Supporting Information. As shown in Figure S3 in the Supporting Information, the FTIR spectrum reveals that γ-NaAsSe$_2$ has no significant absorption up to 16 μm or 600 cm$^{-1}$, and therefore the wavelength transparency range of γ-NaAsSe$_2$ is 0.71 to 16μm. Theoretical prediction confirms the large transparency range as seen in Figure S4 in the Supporting Information.

To understand the electronic structure of γ-NaAsSe$_2$, we performed density functional theory (DFT) calculations using the PBE functional.[33] The electronic band structure of γ-NaAsSe$_2$ reveals relatively flat valence bands and a direct bandgap at the C$_2$ point of 1.18 eV, which is consistent with the well-known underestimation of band gaps by semi-local exchange-correlation functionals (Figure 1d). Atom-projected density of states (PDOS) indicate that the valence band comprises almost exclusively selenium *p* states with small contributions from arsenic *s* and *p* states near the valence band maximum which indicates partial ionicity. The mix of *s* and *p* arsenic states and the trigonal pyramidal coordination of the arsenic atoms with the three neighboring selenium atoms suggests that the covalency of the upper valence states is between arsenic *sp*$^3$ orbitals and selenium *p* states. Lower in the valence band, the arsenic contributes more proportionally to the total DOS.

We next investigated the linear optical properties of γ-NaAsSe$_2$. For a monoclinic system, there are 3 different coordinate systems that do not coincide with one another: crystallographic coordinates (*a, b, c*), crystal physics coordinate ($Z_1, Z_2, Z_3$), and principal eigen coordinate ($Z_1^e, Z_2^e, Z_3^e$). The crystallographic coordinates are defined as *a*//[100], *b*//[010] and *c*//[001]; The crystal physics coordinates are defined as $Z_2$//[010], $Z_3$//[001], and $Z_1$//$Z_2 \times Z_3$. In the principal Eigen coordinate, $Z_2^e$ coincides with the *b* axis and the $Z_2$, the other two eigen axes are obtained by rotating the crystal physics axes about $Z_2^e$ such that the dielectric tensor is diagonalized; α denotes the angle between $Z_3$ and $Z_3^e$, and it varies with wavelength. To study the linear optical properties of γ-NaAsSe$_2$, spectroscopic ellipsometry was used to obtain the complex refractive indices from 0.2μm to 1μm on three different orientations of the crystal. The ellipsometry spectra from all three sample orientations were fitted simultaneously to extract the complex refractive indices. DFT calculations were also performed to understand how the electronic structure produces the derived linear optical properties.

The measured and calculated refractive indices are shown in **Figure 2**. The anisotropic refractive index is defined as, $\tilde{n} = n + ik$. In the monoclinic system, the complex refractive indices have off-diagonal terms when expressed in the crystal physics coordinate. The dielectric tensor in this coordinate system ($Z_1, Z_2, Z_3$) is given by:

$$\tilde{\varepsilon} = \tilde{n}^2 = \begin{pmatrix} \tilde{n}_{11} & 0 & \tilde{n}_{13} \\ 0 & \tilde{n}_{11} & 0 \\ \tilde{n}_{13} & 0 & \tilde{n}_{33} \end{pmatrix}^2. \qquad (1)$$

The *n* and *k* values shown in Figure 2 exhibit excellent qualitative agreement and a quantitative agreement that is typically considered acceptable in such comparisons, considering that DFT is not an excited state theory. For example, in both experiments and the DFT, we find a larger index in the [100] direction than in any other direction. We also find that the value of the off-diagonal element $\tilde{n}_{13}$ is very small and does not affect the fitting of other elements significantly. This can be attributed to the fact that the unit cell angle β is nearly 90°. In other words, the crystal physics axes and the principal eigen axes almost align with each other, and the crystal is practically optically orthorhombic. To confirm this hypothesis, the crystal was investigated with a polarizing microscope, similar to the method reported by Haertle et al.[34] The crystal was rotated about its [010] axis from the [001] axis until the measured intensity is minimal between a pair of cross polarizers. This angle, α, is the rotation angle between the crystal physics axis, $Z_3$ and the principal eigen axis, $Z_3^e$. From these measurements, it was found that $|α| \leq 1.15°$.

Since the NLO properties of interest are at energies below the bandgap, the refractive indices from 0.7μm to 1μm were fitted to the Cauchy equation $n = A + \frac{B}{\lambda^2} + \frac{C}{\lambda^4} + \frac{D}{\lambda^6}$ and extrapolated to 2μm, assuming the dispersion of $n$ is small at lower energies. The parameters of the Cauchy equations are shown in **Table 1**. These values were used for obtaining the second order NLO coefficients.

**Table 1**. The parameters of Cauchy equations for the linear optical properties of γ-NaAsSe$_2$ in the crystal physics coordinate system, ($Z_1, Z_2, Z_3$), from 0.7μm to 1μm.

|          | A      | B         | C           | D        |
|----------|--------|-----------|-------------|----------|
| $n_{11}$ | 2.845  | 0.04496   | -0.02974    | 0.03112  |
| $n_{22}$ | 2.277  | 0.2022    | -0.1249     | 0.03874  |
| $n_{33}$ | 2.375  | -0.004139 | 0.003537    | 0.01321  |
| $n_{13}$ | 0.2122 | 0.001125  | -8.732e-08  | 0.003614 |

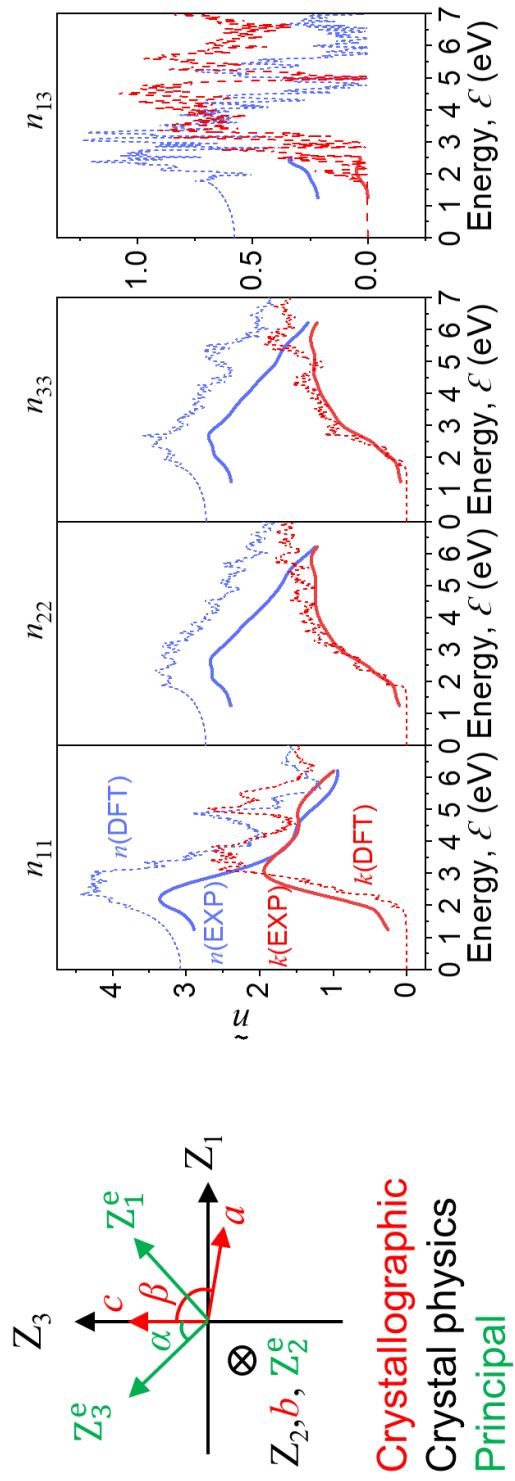

**Figure 2.** (a) 3 coordinate systems for monoclinic materials. (b) The experimental and calculated complex refractive indices of $\gamma$-NaAsSe$_2$ in the visible range. The blue and red curves are n and k, respectively.

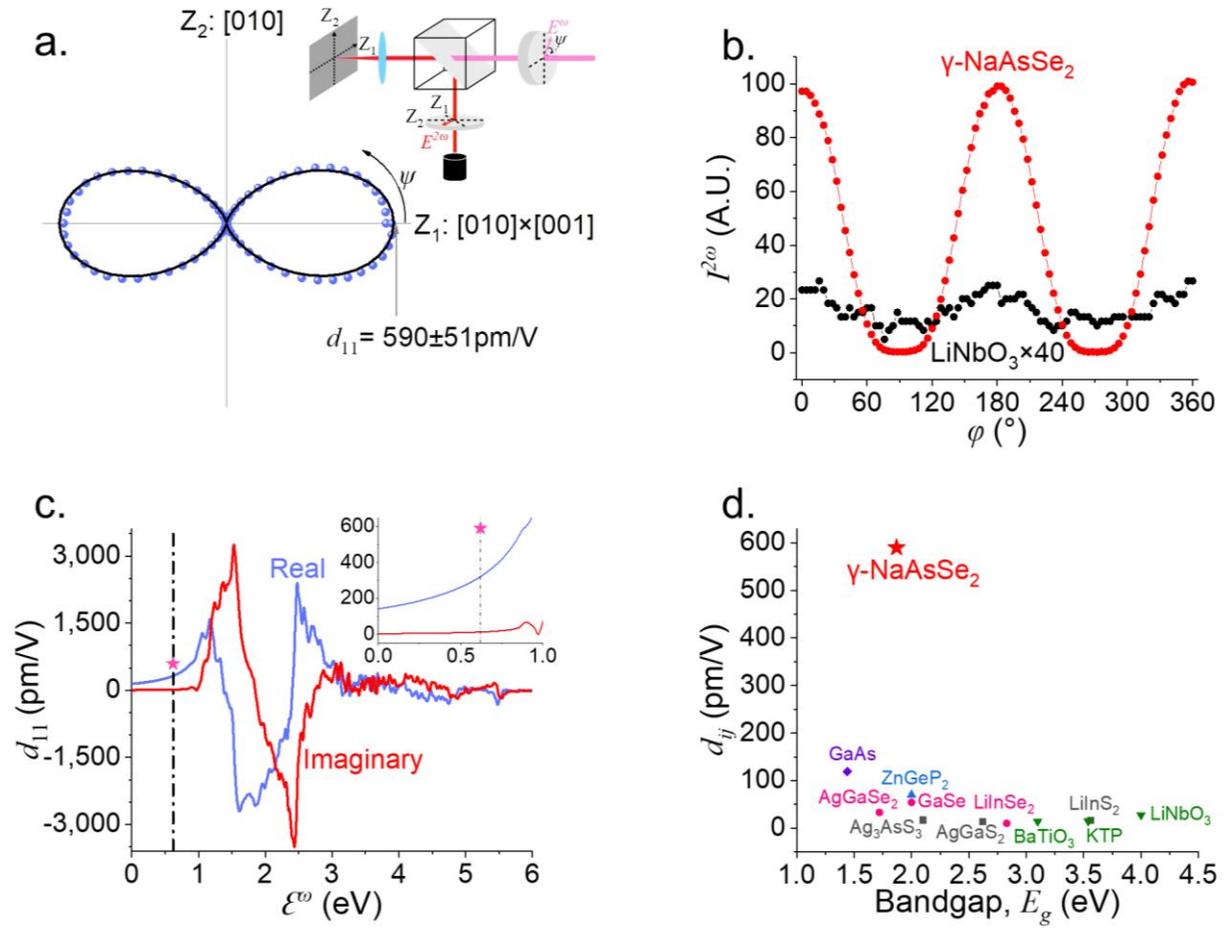

**Figure 3**. (a) Polar plot of p-polarized SHG intensities generated from single crystal γ-NaAsSe$_2$. The black solid line is the simulated SHG values. The inset shows the experimental setup in normal reflection geometry. (b) Comparison of the SHG intensities of γ-NaAsSe$_2$ and LiNbO$_3$, normalized to the same incident power. (c) Calculated $d_{11}$ vs energy. The pink star indicates the experimental value of $d_{11}$. The inset is an enlargement from 0 to 1eV for clarity. (d) Comparison of the highest SHG coefficient and bandgap of some well-known NLO crystals.

We next probe the second order nonlinear optical susceptibility of γ-NaAsSe$_2$. In the process of second harmonic generation (SHG), two photons of frequency $\omega$ combine to form one photon of frequency $2\omega$ ($\omega+\omega=2\omega$) through a nonlinear optical medium. The relationship between the nonlinear polarization and the electric field is $P_{i,\,2\omega} \propto d_{ijk} E_{j,\omega} E_{k,\omega}$, where $d_{ijk}$ is the second-order optical susceptibility and $E$ is the incoming electric field. The fundamental wavelength was chosen to be 2μm such that the SHG wavelength was 1μm, which is within the bandgap, and hence non-resonant and involving only virtual transitions; this minimizes the optical absorption loss. The second harmonic effect was then studied using SHG polarimetry as shown in **Figure 3**a.

The experiment was performed in normal reflection geometry, where the SHG intensities were measured as a function of polarization direction ($\psi$) of the incident field. The $d$ tensor in crystal physics axes, ($Z_1, Z_2, Z_3$), for the point group $m$ in Voigt notation is given as:

$$d = \begin{pmatrix} d_{11} & d_{12} & d_{13} & 0 & d_{15} & 0 \\ 0 & 0 & 0 & d_{24} & 0 & d_{26} \\ d_{31} & d_{32} & d_{33} & 0 & d_{35} & 0 \end{pmatrix}, \qquad (2)$$

The chosen sample geometry allows us to probe the $d_{11}$ and $d_{12}$ coefficients when the analyzer is set to be parallel to the $Z_1$ direction. Figure 3a depicts the measured polar plot fitted to an analytical model based on point group $m$:

$$I_\parallel^{2\omega} \propto \left|E_\parallel^{2\omega}\right|^2 \propto \left(d_{11}t_{\omega,\parallel}{}^2\cos^2\psi + d_{12}t_{\omega,\perp}{}^2\sin^2\psi\right), \qquad (3)$$

where $t_{\omega,\parallel}$ and $t_{\omega,\perp}$ are the Fresnel transmission coefficients for the $p$-polarized and $s$-polarized light, respectively. The absolute magnitudes of the $d$ coefficients can be obtained with respect to an $x$-cut LiNbO$_3$ whose backside is wedged and polished to eliminate the contribution from the back surface. The $d_{33}$ of LiNbO$_3$ is ~ 18 pm V$^{-1}$ at 2μm using Miller's rule.[35,36] By comparing the SHG intensities of γ-NaAsSe$_2$ and LiNbO$_3$ at $\psi = 0°$ shown in Figure 3b, one can determine $d_{11}$=590±51 pm V$^{-1}$ using Equation (S1) and (S2) in the Supporting Information.[37,38] The giant magnitude of $d_{11}$ dominates both the polar plots and the ratio of $d_{11}/d_{12}$; therefore, even a small misalignment leads to the $d_{11}$ "leaking into" the $d_{12}$ value. Hence, only an upper limit for $d_{12}$ can be determined. The ratio of $d_{11}/d_{12}$ is found to be greater than 11, and hence $d_{12} < \sim 54$ pm V$^{-1}$.

DFT calculations were performed to understand the origin of the larger linear and nonlinear optical response in the [100] direction than any other direction. Figure 3c depicts the calculated complex $d_{11}$ of γ-NaAsSe$_2$ compared with the experimental value. From the imaginary components of the linear and nonlinear response spectra, the origin of these much larger responses is attributable to energies associated with low energy valence to conduction band excitations. The high DOS in the valence band and relatively flat bands from the selenium $p$ states provide ample excitation routes, but this is true of all directions. The likely origin then is attributed to the arsenic-selenium chains extending along the [100] direction. It may be tempting to associate the somewhat flatter conduction bands in the $a$-direction, but the band contributions to the overall response are not easy to disambiguate until tools are developed for mapping electron velocity matrix elements onto relevant crystal features.[39]

Figure 3d highlights the highest non-resonant SHG coefficient vs bandgap for some well-known NLO crystals. It clearly shows the general trend that crystals with large bandgaps typically exhibit smaller SHG coefficients. However, $d_{11}$ of γ-NaAsSe$_2$ exceeds the values in all the benchmark NLO crystals. It is worth noting that it surpasses the $d_{36} \approx 33$ pm V$^{-1}$ of AgGaSe$_2$ by eighteen times, though its bandgap is comparable with that of AgGaSe$_2$.[40]

It is interesting to compare the SHG conversion efficiency of γ-NaAsSe$_2$ to conventional NLO materials LiNbO$_3$, AgGaSe$_2$ and ZnGeP$_2$ at various wavelengths. The SHG intensity is proportional to $d^2$ and $l^2$, where $d$ is the SHG coefficient and $l$ is the distance that the light travels.

When not phase-matched, the maximum SHG intensity is generated after traveling a distance of one coherence length, $l_c$.[41] The coherence length is defined as:

$$l_c = \frac{\lambda_\omega}{2(n_{2\omega}-n_\omega)}, \tag{4}$$

in which $n_{2\omega}$ and $n_\omega$ are the refractive indices of the SHG and the fundamental beams. The coherence lengths for the largest SHG coefficients of γ-NaAsSe$_2$ and other NLO materials in the 1.5-2.5μm spectral range are shown in **Figure 4**a. The coherence length of γ-NaAsSe$_2$ is much larger compared to other NLO materials and continues increasing in the infrared range. The promising coherence length will allow us to directly assess the $d_{11}$ coefficient. The SHG conversion efficiency, normalized by $I_\omega^2$ and $l_c$, can be calculated using the following equation:

$$\frac{I_{2\omega}}{I_\omega^2} = \frac{2\omega^2}{\varepsilon_0 c^3} \frac{d^2}{n_{2\omega}n_\omega^2} \frac{1}{l_c} \int_0^{l_c} l^2 \text{sinc}^2\left(\frac{\Delta k l}{2}\right) dl. \tag{5}$$

where $\varepsilon_0$ is the vacuum permittivity and $c$ is the speed of light, and $\Delta k$ is the difference of the wave vectors of the SHG wave and fundamental wave. Figure 4b shows the non-phase-matched SHG conversion efficiency for the largest $d$ coefficients of γ-NaAsSe$_2$ and the commercial NLO materials. At 2μm fundamental wavelength, non-phase-matched SHG from γ-NaAsSe$_2$ is nearly 300 times more efficient than LiNbO$_3$ and 500 times more efficient than AgGaSe$_2$. When compared with ZnGeP$_2$, the SHG conversion efficiency of γ-NaAsSe$_2$ is three orders of magnitude stronger. Combined with the chain-like characteristics which allow easy exfoliation, as well as solubility in strong polar solvents,[27] makes γ-NaAsSe$_2$ an excellent material for developing NLO thin films. The combination of both large $d_{11}$ coefficient and coherence length makes this material an outstanding candidate for exploring orientation patterned quasi-phase-matching approaches.[42]

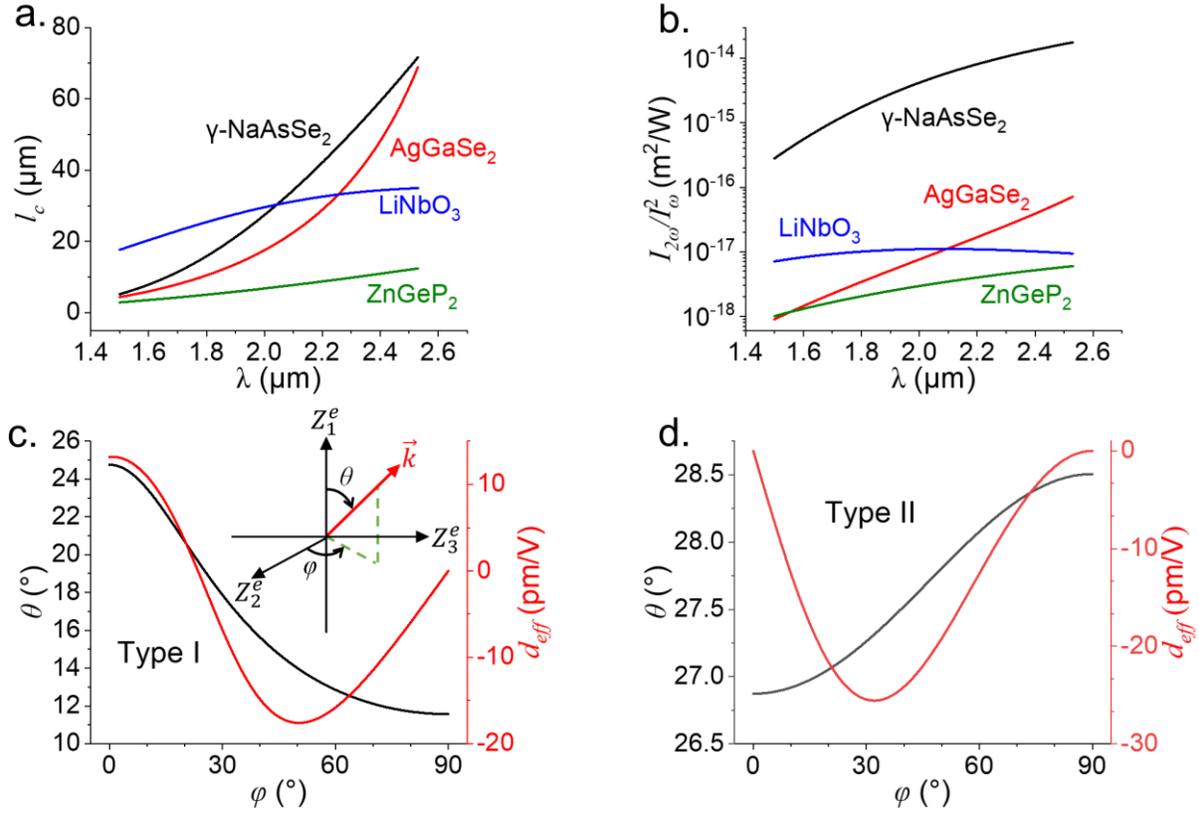

**Figure 4**. Comparison of coherence lengths (a) and SHG conversion efficiency (b) between γ-NaAsSe$_2$ and conventional NLO materials. Type I (c) and Type II (d) phase matching angles (black) and $d_{eff}$ (red) at 2μm fundamental wavelength. Inset in (c) shows the definition of the phase matching angles $\theta$ and $\varphi$.

For applications which demand a high-power nonlinear conversion, the phase matching condition needs to be satisfied in order to eliminate restrictions on the crystal size used and to achieve the most efficient SHG.[43] Under this condition, $\Delta k$ in Equation (5) equals 0; this implies that the refractive indices at ω and 2ω frequency should be equal. Figure 4c and Figure 4d show the Type I and Type II phase matching angles and the $d_{eff}$ at 2μm. The phase matching angles, $\theta$ and $\varphi$, are defined with respect to the Eigen direction, ($Z_1^e$, $Z_2^e$, $Z_3^e$) as shown in the inset of Figure 4c. Using the extrapolated refractive indices from Table 1 and the method reported by Yao and Fahlen,[44] we can then calculate $\theta$ and $\varphi$ at 2μm fundamental wavelength.[44] Based on the calculation, γ-NaAsSe$_2$ can be both Type I and Type II phase matched. The effective $d$ coefficient, $d_{eff}$, can be estimated with the experimental measured $d_{11}$ and the other $d$ coefficients from DFT (Figure S5 and S6, Supporting Information). At 2μm fundamental wavelength, the maximum $d_{eff}$ is -18 pm V$^{-1}$ and -26 pm V$^{-1}$ for Type I and Type II phase matching, respectively. These are comparable to the current commercial crystals of $d_{eff}$ ~ 13.4 pm/V(AgGaS$_2$) and 26.8pm/V(AgGaSe$_2$). The NLO crystals are typically professionally polished to optical grade and their surfaces are coated with anti-reflection coatings to prevent laser damage. Even without these steps, the laser-induced surface damage threshold (LISDT) of a cleaved γ-NaAsSe$_2$ surface (without polishing or coating) to ~25ps pulses at 1.064μm is comparable to that of commercial polished and coated AgGaS$_2$ crystals (see the Supporting Information).

γ-NaAsSe$_2$ single crystals exhibit a giant second order nonlinearity with a remarkable optical SHG coefficient of $d_{11}$=590 pm V$^{-1}$ measured at 2 μm wavelength. This is the highest known non-resonant coefficient (see Figure 3d) for comparable optical bandgaps. For non-phase-matched SHG response generated in one coherence length, γ-NaAsSe$_2$ is two orders of magnitude more efficient than that of the conventional NLO materials at 2μm fundamental wavelength, making it a highly promising candidate to explore towards orientation-patterned quasi-phase-matched devices.[42] In addition, it can also achieve both Type I and Type II phase matching with maximum $|d_{eff,\text{I}}| \approx 18$ pm V$^{-1}$ and $|d_{eff,\text{II}}| \approx 26$ pm V$^{-1}$ at 2μm fundamental wavelength, which are comparable to the current commercial crystals of AgGaS$_2$ and AgGaSe$_2$. These promising optical properties make it a potential candidate to explore large crystals for bulk and quasi-phase-matched high infrared power generation in laser systems.

**Experimental Section**

*Starting materials*: All manipulations were performed under dry nitrogen atmosphere in a glove box. Commercially available chemicals potassium sodium (Na, Sigma Aldrich, 99.5%), arsenic (As, Alfa Aesar, 99.9%) and selenium (Se, American elements, 99.999%) were used without further purification. $Na_2Se$ were prepared by modified literature procedure by reacting the alkali metals and selenium in liquid ammonia.[45] ***Warning: Elemental arsenic is highly toxic which must always be weighed out in the glovebox and precautions must be taken in preparing these samples.***

*Synthesis of the title compounds γ-NaAsSe₂*: Single crystals > 2mm of γ-NaASe$_2$ were grown by the combination of 0.732g Na$_2$Se (5.85 mmol), 0.878g As (11.72 mmol) and 1.388g (17.57 mmol) which were thoroughly ground in the glovebox and loaded in a separate carbon coated fused-silica tube (13 mm OD). Carbon coated silica tubes were used to prevent tube attack from the alkali metal. The tube was then flame sealed under vacuum (~3 x $10^{-3}$ mbar) and inserted in a single zone programmable vertical tube furnace. For the reaction in the vertical furnace the temperature profile used was increasing to 500 °C in 12 h, annealed for 72 h and cooled to 350 °C over 120 h at which point the furnace was turned off. The phase purity of the sample was confirmed using powder x-ray diffraction (XRD) analysis. Single XRD was performed on a crystal with dimensions 0.15 x 0.14 x 0.01 mm$^3$ mounted on a glass fiber with epoxy for structure determination. A summary of the crystal data and refinement is provided in Table S1 in the Supporting Information. Final atomic coordinates and isotropic displacement (Uiso) are listed in Table S2 in the Supporting Information.

*Spectroscopic Ellipsometry*: The spectroscopic ellipsometry was performed using a Woollam M-2000F focused beam spectroscopic ellipsometer on three different orientations of the crystal. The orientations are: 1. [001]// laboratory z, [010]// laboratory x; 2. [001] // laboratory z, [010] // laboratory y and 3. [010]// laboratory z, [001]// laboratory x. The collected ellipsometric spectra were simultaneously fitted to Tauc-Lorentz oscillators and spline function for the diagonal terms and off-diagonal term, respectively. The parameters of the Tauc-Lorentz oscillators include an amplitude $A_m$, full width at half-maximum (FWHM) $B_m$, energy center $E_{0,m}$ and a Tauc gap $E_{g,m}$ (Table S3, Supporting Information).

*SHG measurements*: The fundamental beam of 2μm generated from a Spectra-Physics Ti: sapphire pumped OPA-800C (100fs, 1kHz) was linearly polarized and rotated by an angle of ψ using a half waveplate and focused on the sample surface. The reflected second harmonic beam was filtered out by a dichroic mirror and detected by a photo-multiplier tube after decomposed into a *p*-polarized and *s*-polarized light by an analyzer.

*DFT calculation*: All Density Functional Theory (DFT) calculations were performed using the PBE functional.[33] ABINIT version 9.2.2 was used for linear and non-linear optical properties calculations.[46] Simulation preparation and post-processing are performed with the Atomic Simulation Environment (ASE) version 3.19.1. [47]

Single point calculations were performed with ABINIT preceding the optical properties calculations within single particle approximation using the ABINIT utility, Optic. The Briliouin zone was sampled with a 4×8×4 *k*-point grid. The plane wave cutoff was 1200 eV. The electronic solver convergence criterion was such that no energy eigenvalue changed by more than $10^{-6}$ eV between steps. Pseudopotentials from the standard accuracy, scalar relativistic set of the optimized norm-conserving Vanderbilt pseudopotentials (ONCV) version 0.4.0 were employed.[48] The number of empty bands was increased until the highest empty band was 20 eV above the valence band maximum which corresponds to 232 empty bands. Optics calculations within the independent particle approximation were performed using 50 meV broadening and with scissor shifts of 0.48 eV to compensate for the difference between the DFT fundamental gap and the experimentally measured bandgap.[49]

**Supporting Information**
Supporting Information is available from the Wiley Online Library or from the author.

**Acknowledgments**
J.H., A. K. I, S. S., M.G.K, and V.G. acknowledge the Air Force Office of Scientific Research Grant number FA9550-18-S-0003. M.J.W and J. M. R. were supported by the National Science Foundation's (NSF) MRSEC program (DMR-1720139) at the Materials Research Center of Northwestern University. Scientific discussions and advice from Gary Cook, Carl M. Liebig, Ryan K. Feaver, and Sean A. McDaniel from the AFRL are gratefully acknowledged. J.H. and A. K. I contributed equally to this work.

# Supporting Information

# Giant Non-resonant Infrared Second Order Nonlinearity in γ-NaAsSe$_2$

*Jingyang He,[†] Abishek K. Iyer,[†] Michael J. Waters, Sumanta Sarkar, James M. Rondinelli, Mercouri G. Kanatzidis,* Venkatraman Gopalan**

† Equal contributions

J. He, Prof. V. Gopalan
Department of Materials Science and Engineering, Pennsylvania State University, University Park, Pennsylvania, 16802, USA
Email: vxg8@psu.edu

Dr. A. Iyer, Dr. S. Sarkar, Prof. M. Kanatzidis
Department of Chemistry, Northwestern University, Evanston, Illinois 60208, USA
Email: m-kanatzidis@northwestern.edu

Dr. M. Waters, Prof. J. Rondinelli
Department of Materials Science and Engineering, Northwestern University, Evanston, Illinois, 60208, USA

**Crystal Growth of γ-NaAsSe$_2$ by Bridgman Method**

In the first step, the polycrystalline phase of γ-NaAsSe$_2$ was synthesized by reacting 0.2415 g Na (10.5 mmol), 0.7492 g As (10 mmol) and 1.5792 g of Se (20 mmol) in a vacuum sealed and carbon coated fused silica tube. The tube was slowly heated to 500 °C over 15 h and homogenized at that temperature for 8 h. This was followed by fast cooling to 250 °C over 8h and annealing for 72 h. Finally, the tube was quenched in ice cooled water. Subsequently, the as-synthesized ingot was transferred into a conical-tip and carbon coated quartz tube with an inner diameter of 10 mm and flame-sealed under vacuum (~ 3 x 10$^{-3}$ mbar). The sealed ampoule was placed inside a vertical two-zone Bridgman furnace to grow large single crystals of γ-NaAsSe$_2$. The ampoule was first kept in the upper hot zone of the Bridgman furnace at 550 °C for 12 h to ensure thorough melting of the ingot. Next, the ampoule was translated from the upper hot zone to the lower cold zone at a descending speed of 0.5 mm h$^{-1}$. The temperature of the cold zone temperature was set at 300 °C. After the crystallization was complete, the ampoule was kept at 300 ° C inside the cold zone for another 72 h and finally quenched in ice-cold water.

**Crystal Characterization**

**Crystal Structure Characterization**

X-ray powder diffraction patterns were collected on a Rigaku Miniflex600 diffractometer using Cu-Kα1 radiation ($\lambda$ = 0.154593 nm) equipped with a high-speed silicon strip detector. Finely powdered samples were measured on a flat zero background Si sample. The experimental patterns were compared to simulated patterns based on the experimental CIF files. The crystal structure was determined by single-crystal X-ray diffraction methods. Plate like crystals were chosen for the single-crystal X-ray diffraction study. Data collections were done at 293 K using a STOE imaging plate diffraction system (IPDS-II) with graphite-monochromated Mo Kα radiation operating at 50 kV and 40 mA with a 34 cm image plate. Individual frames were collected with scan widths of 1.0° in $\omega$ and 15 min exposure time. The X-AREA, X-RED, and X-SHAPE software packages were used for data extraction and integration and to apply analytical absorption corrections. Direct methods and full-matrix least-squares refinement against $F2$ were performed with the SHELXTL/2018 package.

**Diffuse reflectance Spectroscopy**

Optical diffuse reflectance measurements were performed at room temperature using a PerkinElmer LAMBDA 1050 UV/Vis spectrophotometer in the range of 1500 nm-250 nm. The instrument is equipped with an integrating sphere detector and controlled by a computer. $BaSO_4$ was used as a 100% reflectance standard. The samples were prepared by grinding the crystals to a powder and spreading it on a compacted surface of the powdered standard material, preloaded into a sample holder. The reflectance versus wavelength data generated were used to estimate the bandgap of the materials by converting reflectance to absorption data using the Kubelka-Munk equation and applying the extrapolation method.[1]

**Differential Thermal Analysis**

Differential thermal analysis (DTA) was performed on the sample using a Shimadzu DTA-50 thermogravimetric analyzer to determine its thermal stability. Approximately 50 mg of the ground crystalline material was flame-sealed in a fused silica DTA tube and evacuated to ~ $10^{-3}$ mbar. A similarly sealed ampule of ~50 mg of $Al_2O_3$ was used as a reference. Both the Sample and reference were heated to ∼ 600 °C at 5 °C min$^{-1}$ and then cooled to room temperature at the same rate.

Differential thermal analysis on the as-synthesized sample of γ-NaAsSe$_2$ showed the sample to be incongruently melting and converting to the δ-NaAsSe$_2$(Figure S1a, b). The melting point of γ-NaAsSe$_2$ was found to be 444 °C ($T_m$) and the crystallization temperature ($T_c$) 420°C. Variable temperature powder XRD was used to confirm this polymorphic transformation (Figure S1c, d). Figure S1d shows that once the melting point of the material is reached, the cooling of the sample results in the polymorphic δ-NaAsSe$_2$. This polymorphic transition was not observed only under the Bridgeman conditions suggesting that slow cooling rates are essential for the crystal growth of γ-NaAsSe$_2$.

**Attenuated Total Reflection Fourier-Transform Infrared Spectroscopy (ATR-FTIR)**

FTIR data was collected on freshly prepared γ-NaAsSe$_2$ at room temperature on the Bruker Tensor 37 FTIR equipped with mid-IR detector and KBr beam splitter. The spectrum was collected in ATR mode in the rage of 2.5 to 16μm (4000 to 600 cm$^{-1}$). The data was averaged over 16 scans. The OPUS software was used for the data acquisition.

**Variable temperature powder X-ray diffraction**

In a nitrogen filled glovebox, the same mass of reagents used for bulk synthesis were combined in an agate mortar and pestle and ground for 10 min. The powder was loaded into a 0.5 mm fused silica capillary and flame sealed under 3 × 10$^{-3}$ mbar. Variable temperature diffraction patterns were collected on a STOE STADI – MP diffractometer. The capillary was placed in the furnace attachment which has temperature stability of 0.1°C. This diffractometer uses an asymmetric curved Germanium monochromator to select the Mo Kα₁ line (λ = 0.70930 Å) and has a one-dimensional silicon strip detector (MYTHEN2 1K, DECTRIS). The X-ray generator operates at 50 kV and 40 mA. Prior to measurement, calibration was performed using a NIST silicon standard (640d). The heating profile used can be seen in Figure S1c.

**Calculation of SHG coefficient for γ-NaAsSe$_2$**

The reflected SHG intensity of the single surface of a non-absorbing material can be expressed as[2]

$$I_{2\omega} = (I_\omega)^2 (dt_{\omega,||})^2 \Omega, \tag{S1}$$

where $\Omega = \left(\frac{2}{c\varepsilon_0}\right)\left(\frac{r_{2\omega,||}}{(1-n_\omega)(n_\omega+n_{2\omega})}\right)^2$.

The absolute value of $d_{11}$ of γ-NaAsSe$_2$ can be calculated using the following equation:

$$\left|d_{11}^{\gamma-NaAsSe_2}\right| = \left|d_{33}^{LiNbO_3}\left(\frac{t_{\omega,||}^{LiNbO_3}}{t_{\omega,||}^{\gamma-NaAsSe_2}}\right)^2 \left(\frac{I_\omega^{LiNbO_3}}{I_\omega^{\gamma-NaAsSe_2}}\right)\sqrt{\frac{I_{2\omega}^{\gamma-NaAsSe_2}\Omega^{LiNbO_3}}{I_{2\omega}^{LiNbO_3}\Omega^{\gamma-NaAsSe_2}}}\right| \tag{S2}$$

Here, $t_{\omega,||}$, the Fresnel transmission coefficient at 2ω, is defined as $t_{\omega,||} = \frac{2}{1+n_\omega}$. The Fresnel reflection coefficient, $r_{2\omega,||}$, is defined as $r_{2\omega,||} = \frac{n_{2\omega}-1}{n_{2\omega}+1}$.

**Laser induced surface damage threshold (LISDT) measurements**

The Laser induced surface damage threshold (LISDT) of γ-NaAsSe$_2$ was measured with both femtosecond pulses and picosecond pulses: a 1.55μm laser beam (100fs, 1kHz) generated from Spectra-Physics Ti: sapphire pumped OPA-800C and a 1.064μm laser beam (27ps, 10Hz) generated from an optical parametric generator/amplifier (OPG/OPA) system pumped by a Nd: Yag laser. For both measurements, the beam was focused on a freshly cleaved surface of the crystal without polishing and coating. The optical microscope (Zeiss Axioplan 2) with 10x magnification was used to examine the surface of the sample after each trial of test. The power was gradually increased until apparent damage was observed on the surface. The LISDT value of a cleaved, unpolished and uncoated γ-NaAsSe$_2$ surface was found to be 0.34 GW cm$^{-2}$ and 0.24 GW cm$^{-2}$ for the femtosecond pulses (at 1.55μm) and picosecond pulses (at 1.064 μm) respectively. In contrast, the LISDT of a polished and coated AgGaS$_2$ crystal is 0.7 GW cm$^{-2}$ measured with 25ps pulses, which is only a factor of 2 better. Thus, we expect that when the surface of γ-NaAsSe$_2$ is commercially polished and coated with an antireflection optical coating, its LISDT will significantly improve.

**Supplementary Figure**

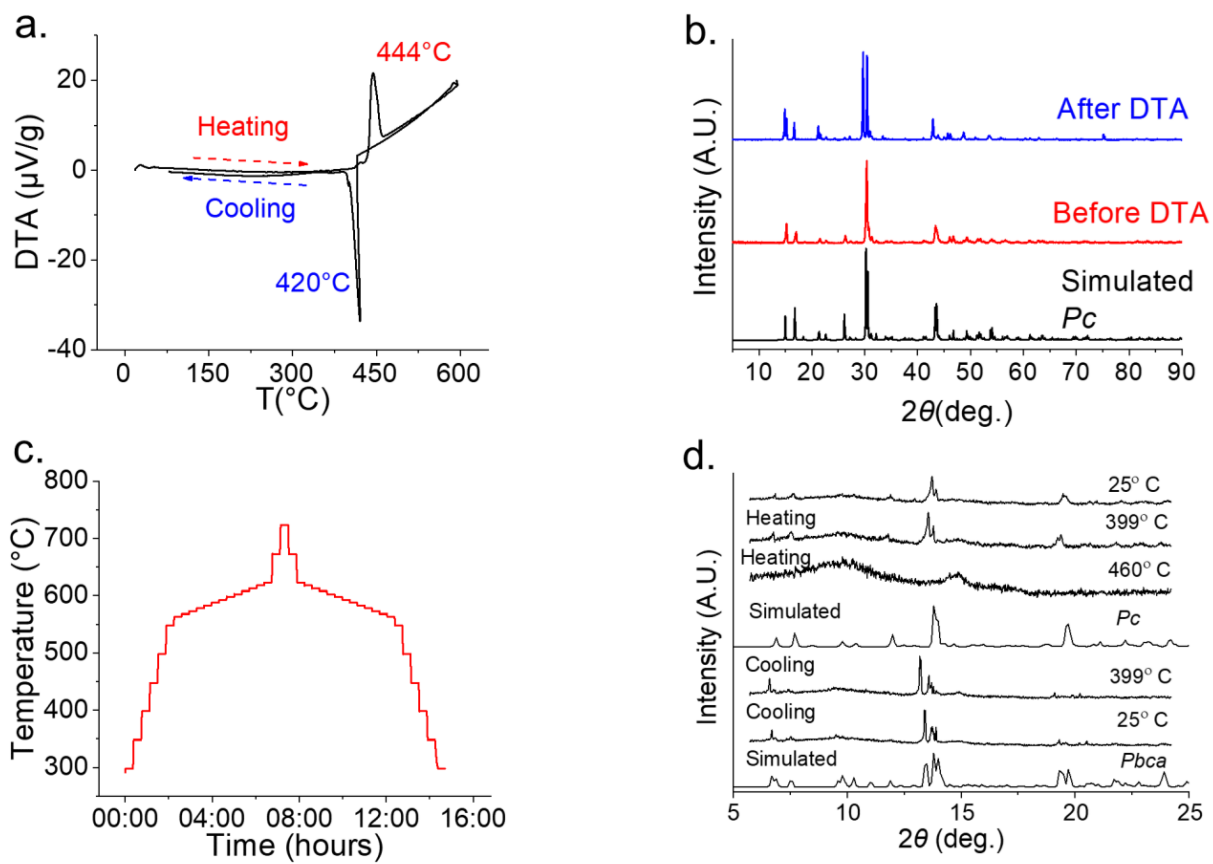

**Figure S1.** (a) Differential thermal analysis for γ-NaAsSe$_2$ showing the melting point and crystallizing point. (b) the comparison of the powder XRD pattern before and after DTA measurement. (c) temperature profile during the variable temperature powder XRD measurement. (d) Variable temperature powder XRD of γ-NaAsSe$_2$.

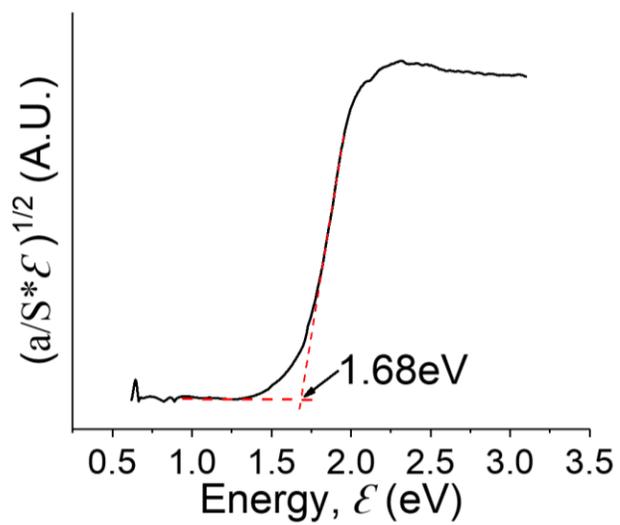

**Figure S2.** Tauc analysis of indirect bandgap for γ-NaAsSe$_2$.

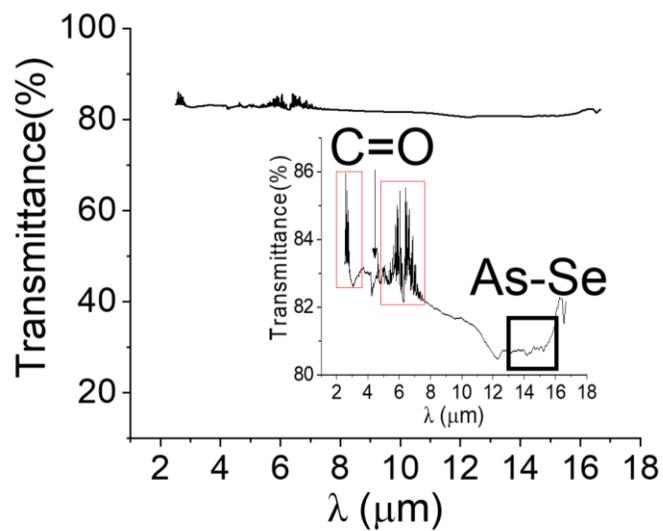

**Figure S3.** FTIR spectrum collected on freshly prepared powder of γ-NaAsSe$_2$ shows no significant absorption up to 16 μm or 600 cm$^{-1}$. The inset shows a weak absorption between 15.9μm (630 cm$^{-1}$) to 13.5μm (740 cm$^{-1}$) which can be attributed to As-Se vibrational stretching, an absorption from CO$_2$ at 4.2μm (2356 cm$^{-1}$) attributed to C=O stretching[3] and the noise in the red boxes corresponds to the change in grating within the instrument.

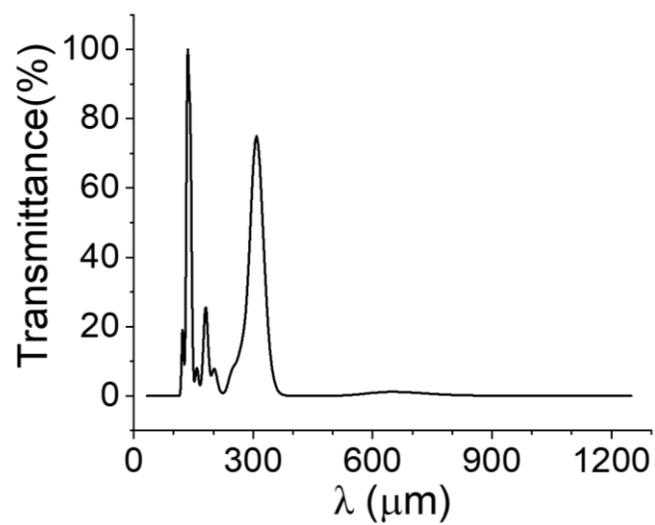

**Figure S4.** FTIR spectrum calculated by DFT indicates that the γ-NaAsSe$_2$ is transparent from 33 μm (300 cm$^{-1}$) to 100 μm (100 cm$^{-1}$) and from 390 (25.6 cm$^{-1}$) to 1250 μm (8 cm$^{-1}$).

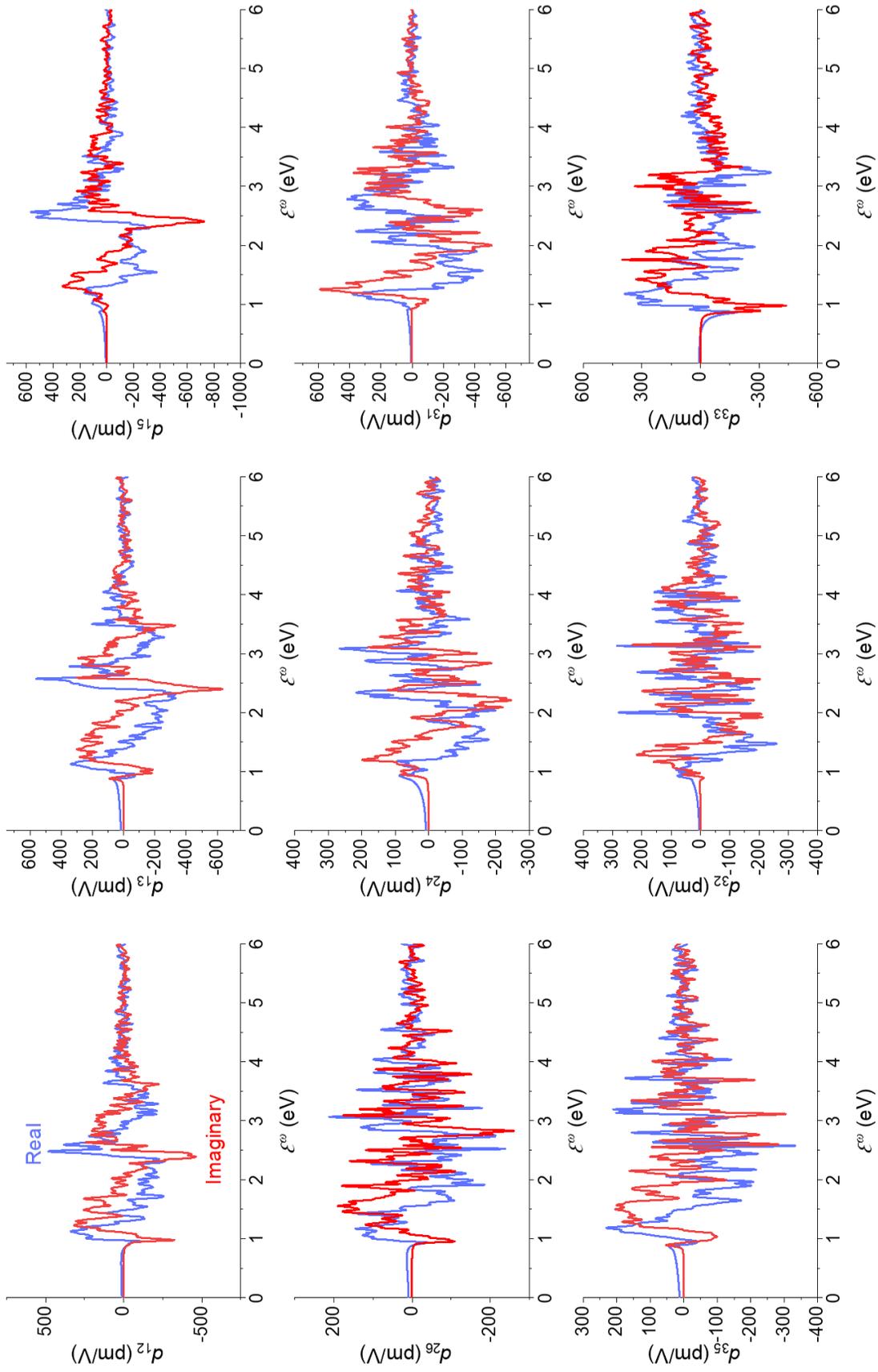

**Fig. S5.** Calculated *SHG* coefficients $d_{ij}$ vs fundamental energy.

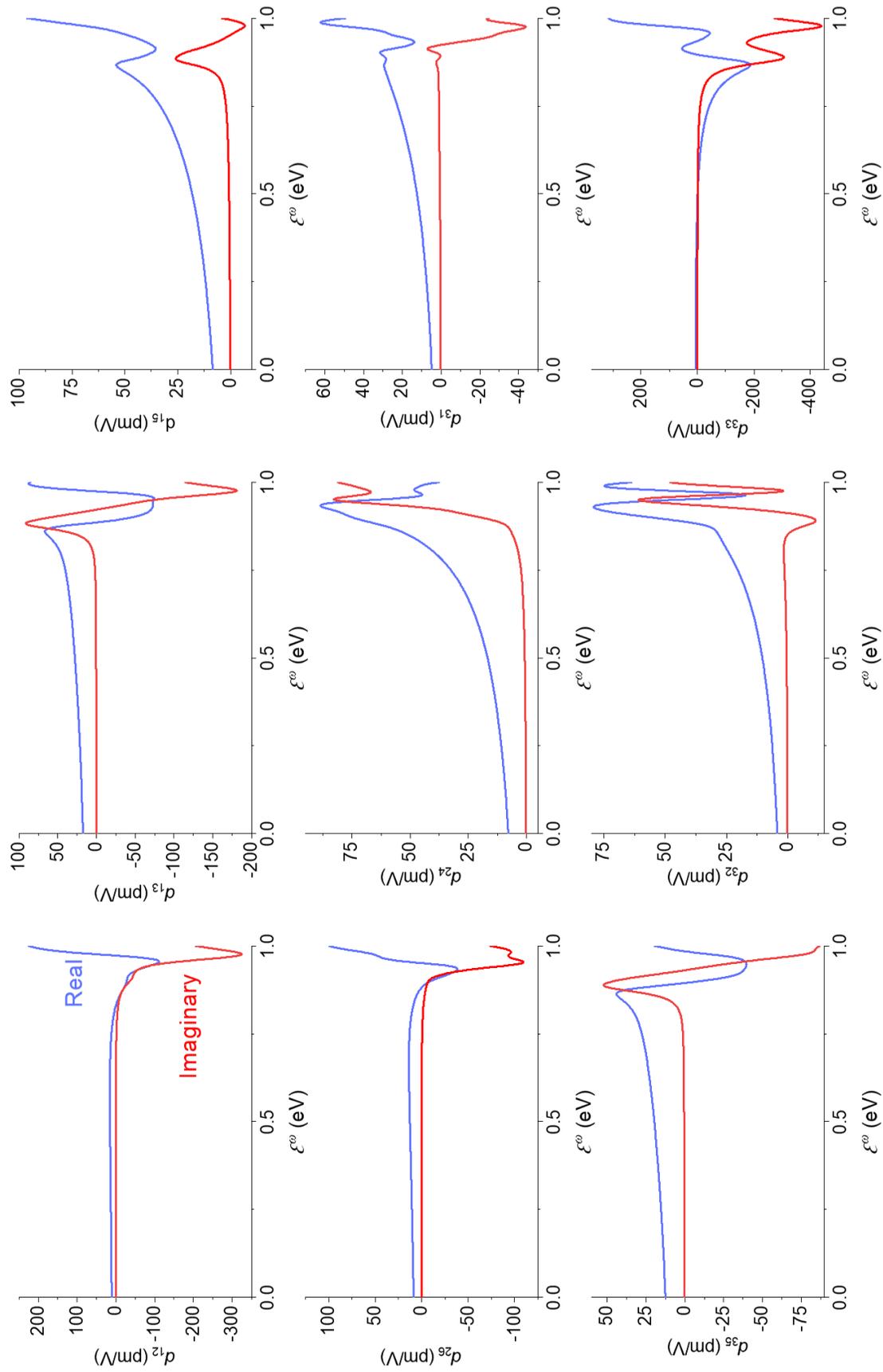

**Fig. S6.** Calculated SHG coefficients $d_{ij}$ vs fundamental energy in the range of 0 to 1 eV.

**Supplementary Table**

Table S1. Crystal data and structure refinement for γ-NaAsSe$_2$ at 293K.

| | |
|---|---|
| Empirical formula | NaAsSe$_2$ |
| Formula weight | 1023.32 |
| Temperature | 293(2) K |
| Wavelength | 0.71073 Å |
| Crystal system | Monoclinic |
| Space group | *Pc* |
| Unit cell dimensions | $a$ = 11.726(2) Å, α = 90°<br>$b$ = 5.9329(12) Å, β = 90.39(3)°<br>$c$ = 11.866(2) Å, γ = 90° |
| Volume | 825.5(3) Å$^3$ |
| Z | 2 |
| Density (calculated) | 4.117 g/cm$^3$ |
| Absorption coefficient | 25.735 mm$^{-1}$ |
| F(000) | 896 |
| Crystal size | 0.15 x 0.14 x 0.01 mm$^3$ |
| θ range for data collection | 3.434 to 33.363° |
| Index ranges | -18<=h<=16, -9<=k<=9, -18<=l<=18 |
| Reflections collected | 9712 |
| Independent reflections | 5665 [R$_{int}$ = 0.0434] |
| Completeness to θ = 25.242° | 99.9% |
| Refinement method | Full-matrix least-squares on F$^2$ |
| Data / restraints / parameters | 5665 / 2 / 146 |
| Goodness-of-fit | 1.025 |
| Final R indices [I > 2σ(I)] | R$_{obs}$ = 0.0511, wR$_{obs}$ = 0.1261 |
| R indices [all data] | R$_{all}$ = 0.0891, wR$_{all}$ = 0.1435 |
| Extinction coefficient | 0.0021(4) |
| Largest diff. peak and hole | 1.813 and -1.590 e·Å$^{-3}$ |

R = Σ||F$_o$|-|F$_c$|| / Σ|F$_o$|, wR = {Σ[w(|F$_o$|$^2$ - |F$_c$|$^2$)$^2$] / Σ[w(|F$_o$|$^4$)]}$^{1/2}$ and w=1/[σ$^2$(Fo$^2$)+(0.1218P)$^2$] where P=(Fo$^2$+2Fc$^2$)/3

**Table S2**. Atomic coordinates (×10$^4$) and equivalent isotropic displacement parameters (Å$^2$×10$^3$) for γ-NaAsSe$_2$ at 293K with estimated standard deviations in parentheses.

| Label | x | y | z | Occupancy | U$_{eq}$* |
|---|---|---|---|---|---|
| Na(01) | 205(8) | 5195(15) | 2352(7) | 1 | 32(2) |
| Na(02) | 2612(9) | 4741(16) | 4988(8) | 1 | 39(2) |
| Na(03) | 5193(8) | 4923(14) | 2678(7) | 1 | 32(2) |
| Na(04) | 7660(10) | 171(16) | 7568(10) | 1 | 46(3) |
| As(01) | 0 | 939(4) | 1(2) | 1 | 30(1) |
| As(02) | 2513(2) | 687(3) | 2254(2) | 1 | 28(1) |
| As(03) | 5026(3) | 78(3) | 533(2) | 1 | 30(1) |
| As(04) | 7505(3) | 4519(4) | 382(2) | 1 | 29(1) |
| Se(01) | 100(3) | 5155(3) | 4861(2) | 1 | 33(1) |
| Se(02) | 389(3) | 245(3) | 1997(2) | 1 | 30(1) |
| Se(03) | 2701(3) | 4588(3) | 2472(2) | 1 | 29(1) |
| Se(04) | 2921(3) | 237(3) | 247(2) | 1 | 30(1) |
| Se(05) | 5105(3) | 97(4) | 7489(2) | 1 | 33(1) |
| Se(06) | 5378(2) | 4056(3) | 146(2) | 1 | 30(1) |
| Se(07) | 7715(3) | 4804(4) | 2338(2) | 1 | 30(1) |
| Se(08) | 7897(3) | 507(3) | 69(2) | 1 | 29(1) |

*U$_{eq}$ is defined as one third of the trace of the orthogonalized U$_{ij}$ tensor.

**Table S3**. The parameters of each Tauc-Lorentz oscillator for the linear optical properties

| m | $A_m^{11}$ | $B_m^{11}$ | $E_{0,m}^{11}$ | $E_{g,m}^{11}$ | $A_m^{22}$ | $B_m^{22}$ | $E_{0,m}^{22}$ | $E_{g,m}^{22}$ | $A_m^{33}$ | $B_m^{33}$ | $E_{0,m}^{33}$ | $E_{g,m}^{33}$ |
|---|---|---|---|---|---|---|---|---|---|---|---|---|
| 1 | 144.2478 | 1.550 | 1.813 | 1.813 | 28.8956 | 1.366 | 2.777 | 1.850 | 16.2442 | 1.063 | 2.844 | 1.797 |
| 2 | 24.0366 | 2.316 | 0.710 | 0.721 | 23.5741 | 2.349 | 3.855 | 1.708 | 14.8376 | 2.330 | 3.952 | 1.045 |
| 3 | 12.5520 | 1.502 | 4.737 | 2.862 | 31.0370 | 2.367 | 4.535 | 3.334 | 3.5961 | 0.641 | 2.179 | 1.404 |
| 4 | 21.0505 | 1.500 | 2.746 | 0.926 | 11.7716 | 1.441 | 5.674 | 3.587 | 1.3250 | 1.143 | 1.379 | 0.588 |
| 5 | - | - | - | - | 3.4840 | 0.583 | 2.186 | 1.320 | 2.6151 | 1.328 | 4.819 | 0.0820 |
| 6 | - | - | - | - | 15.8119 | 2.175 | 0.884 | 1.000 | 134.6440 | 1.555 | 5.218 | 4.638 |